\documentclass[aps, twocolumn,floats, showpacs]{revtex4}
\usepackage{amsmath,amssymb,graphicx}
\usepackage{epsfig}
\usepackage{graphicx}
\usepackage{enumerate}

\newcommand{\beq}{\begin{equation}}
\newcommand{\eeq}{\end{equation}}

\newcommand{\bea}{\begin{eqnarray}}
\newcommand{\eea}{\end{eqnarray}}

\begin{document}
\title{Minimal model of a heat engine: An information theory approach}
\author{Yun Zhou and Dvira Segal}
\affiliation{Chemical Physics Theory Group, Department of Chemistry, University of Toronto,
80 Saint George St. Toronto, Ontario, Canada M5S 3H6}

\begin{abstract}
We construct a generic model for a heat engine using information
theory concepts, attributing irreversible energy dissipation to the
information transmission channels.
Using several forms for the channel capacity, classical and quantum,
we demonstrate that our model recovers both the Carnot principle in
the reversible limit, and the universal maximum power efficiency
expression of nonreversible thermodynamics in the linear response
regime. We expect the model to be immensely useful as a testbed for
studying fundamental topics in thermodynamics.
\end{abstract}


\maketitle



\textit{Introduction.---} The Maxwell's demon puzzle, suggesting a
violation of the second law of thermodynamics, has been exorcised
using Landauer's memory erasure principle \cite{Landauer}. Treating
the demon's intelligence as information has manifested a fundamental
connection
between information and physics, revealing the specific role of memory 
in terms of the second law of thermodynamics \cite{NoriRev}.

The goal of this paper is to suggest a minimal-universal model of a heat engine (demon)
using information theory concepts, and to employ it for
analyzing the operation principles of finite-time thermodynamic machines \cite{FTT}.
In our engine irreversible energy dissipation is attributed to information transfer within
the demon's channels, compensating entropy decrease in
the system. 
The model serves as a testbed for studying fundamental topics in thermodynamics:
The universality of the maximum power efficiency and its relationship to the Carnot efficiency.

A central topic in thermodynamics is the study of the efficiency of thermal
engines, manifesting universal features.
According to the Carnot result the efficiency (work output over heat input)
of a cyclic thermal machine, operating
between two heat baths temperatures $T_C$ and $T_H$ ($T_C<T_H$), is at most
\bea \eta_C=1-\frac{T_C}{T_H}. \eea
The upper limit is obtained for engines that work reversibly. However, as
reversible processes occur infinitesimally slowly, the power produced
is zero.
Operating away from equilibrium, a more practical question is the efficiency at {\it maximum power}, $\eta_M$,
optimizing an engine cycle with respect to its {\it power} rather than efficiency.
For a specific model, Curzon and Ahlborn (CA) derived the maximum power efficiency \cite{Curzon}
\bea
\eta_{CA}&=&1-\sqrt{\frac{T_C}{T_H}}= 1-\sqrt{1-\eta_C}
\nonumber\\
&\approx& \eta_C/2 +\eta_C^2/8 +{\cal O}(\eta_C^3),
\label{eq:etaCA} \eea
relaying on the endoreversible approximation
in which the sole source of irreversibility is due to heat transfer processes \cite{FTT}.
Does (\ref{eq:etaCA}) represent a universal characteristic
of finite-time thermodynamics, or does it depend on the specific model?
In the linear response regime it has been recently proven that the
efficiency at maximum power is upper bounded by
(\ref{eq:etaCA}), which in this regime is exactly half of the Carnot efficiency \cite{broeck}.
The upper limit is reached for a specific class of "strongly coupled" models for which the
energy flux is directly proportional to the work-generating flux.
%
In contrast, in the nonlinear regime general results are missing, and expressions deviating
from the CA efficiency were reported  \cite{Johal,Seifrt}.


\begin{figure}
\hspace{2mm} 
\rotatebox{-90} {\includegraphics[width=38mm]{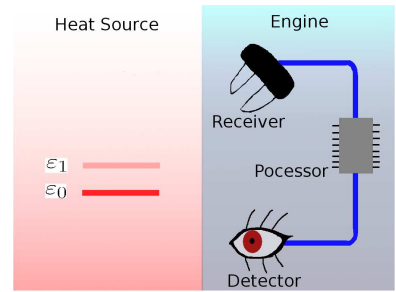}}
\caption{A schematic
representation of the model heat engine, see description in the text.} 
\label{engine}
\end{figure}

In the present model irreversible loss of energy occurs at the
communication channels, transmitting classical information (entropy)
from a heat source ($T_H$) to the engine ($T_C$).
We consider both classical and quantum channels,
encoding classical information in quantum states,
and show that in both cases
Carnot efficiency can be achieved when the device works
reversibly. We also demonstrate that in the linear regime (small temperature differences)
the efficiency at maximum power is $\eta_C/2$.
Using a generic form for the channel capacity we show that an agreement
to higher terms in (\ref{eq:etaCA}) can be obtained for a class of
channels.
Our study indicates  on the fundamental status of the (linear) CA
efficiency in irreversible thermodynamics. We find that
if a model obeys the Carnot limit for a reversible process,
it must also follow, for a nonreversible operation, the universal behavior,
$\eta_M=\eta_C/2$ in the linear response regime, and vice versa.
Moreover, we find that the ratio of $1/2$ between the maximum efficiency and the max-power efficiency holds even
{\it below} the theoretical upper bounds, for imperfect systems.


\textit{Model.---} Our model includes a finite subsystem immersed in a heat
bath and an engine (demon), see Fig. \ref{engine}. In particular, the
subsystem may include two discrete states 0 and 1 of energies
$\varepsilon_0=0$ and $\varepsilon_1=\varepsilon$ respectively.
The subsystem is assumed to be tightly attached to a heat bath at
temperature $T_H$, thus in thermal equilibrium the levels population
follows the Boltzmann distribution
$P_{0}=\frac{1}{1+\exp(-\varepsilon/T_H)}$,
 $P_{1}=\frac{\exp(-\varepsilon/T_{H})}{1+\exp(-\varepsilon/T_{H})}$;
$k_{b}=1$. As the two level system (TLS) is virtually a part of the
heat bath, we refer to the combined object as a \emph{heat source}
hereafter. It is important to note that the TLS is held in
isothermal conditions: it is tightly connected to a large thermal
reservoir at $T_H$. Thus, the engine absorbs energy from the TLS,
yet the TLS does not cool down to zero, as its temperature is kept
fixed. This is analogous to the process of reversible isothermal
expansion of ideal gases. We emphasize that our results do not
depend on the particular choice of the subsystem, and the TLS model
is introduced here as a simple demonstration.

The right half of the plot shows an automatic machine, referred to as
the \emph{engine}.
It consists four components: detector, communication channel, processor and receiver.
Following Shanon general model for a communication system
\cite{EBook}, in a working cycle the detector detects the state of
the heat source, encodes it, and sends this information through the
transmission channel to the processor which will decode it. Based on
this information, the processor sends an instruction to properly set
the receiver, to accept the energy of the subsystem. While a
measurement of the heat source state and receiving its energy
can be done (virtually) without energy consumption \cite{NoriRev},
energy dissipation in the engine is attributed to the
transmission of information in the channels and the resulting
setup of the receiver, corresponding to the Landauer erasure principle \cite{NoriRev}. 

\textit{Analysis of performance.---}
In each working cycle the engine needs to acquire the information on the
state of the finite subsystem. The total amount of information that has to
be absorbed during this process is given by
\bea I= \frac{\bar E}{T_H},
\label{eq:I} \eea
where, e.g., $\bar E =P_0\varepsilon_0 +P_1\varepsilon_1$ is the average
energy of the TLS subsystem.
%
This information has to be transferred through the engine communication channel.
The measure for a communication system is the
channel capacity $I_p(S,N_0)$, describing the maximum number of bits
that a system can communicate reliably per channel use.
Here $S$ and $N_0$ (a function of $T_C$)
are the pulse and noise average energies, respectively \cite{units}.
In order to transfer $I$ on the status of the heat source to the engine
we need to invest energy
\bea
Q\equiv\frac{I}{I_p}S,
\label{eq:Q}
\eea
which cannot be refunded (else it contradicts the maximum capacity theorem),
thus this is the heat dissipated in the engine
during an operation cycle. The engine efficiency is given by the
amount of available working energy over the invested energy. Using
$W=\bar{E}-Q$ for the work attained per cycle, we obtain
\bea
\eta= \frac{W}{\bar E}= \frac{\bar E- Q}{\bar E} = 1- \frac{S}{T_HI_p}.
\label{eq:eta}
\eea
We will optimize the engine cycle (i) with respect to its efficiency, minimizing $Q$, and (ii)
with respect to its power $P= W/n$, where $W=\bar{E}-Q$ is the work attained
per cycle. Here $n=I/I_p$, 
 the number of pulses that are needed to transfer the information $I$ in a working cycle, 
 is proportional to the period of the engine.
Utilizing Eqs. (\ref{eq:I}) and (\ref{eq:Q}) we  get
\bea P=T_H I_p-S.
\label{eq:P}
\eea
Maximizing the power with respect to the pulse power $S$ and substituting the optimal
value $S_M$ into (\ref{eq:eta}), provides the maximum power efficiency
\bea
\eta_M=1-S_M/T_H I_p(S_M,T_C).
\label{eq:etaM}
\eea
We consider next classical and quantum channels, 
derive $\eta$ for the two scenarios mentioned above, and manifest that they  independently follow
a universal behavior.


\textit{Case I: Gaussian Channel.---} Suppose we send information
over a classical continuous memoryless channel of bandwidth $B$ (measured in Hertz),
subjected to an additive white Gaussian noise with a power spectral density $N_0$.
Assuming a signal power $\mathcal S$,
and taking $BN_0$ as the total noise power in a bandwidth $B$,
one obtains the Shanon-Hartley capacity
%
$C=B \ln\left(1+ \frac{\mathcal S}{BN_0}\right)$ \cite{EBook}. 
%
Defining $S=\mathcal S/B$ as the average energy carried by the pulse
and $I_p=C/B$  as a dimensionless capacity \cite{units}, we get
\bea I_p =\ln \left( 1+ \frac{S}{N_0} \right).
\label{eq:Gauss2} \eea
In order to transfer the information $I$ on the status of the
subsystem into the engine we invest
$Q= I\frac{S}{\ln(1+\frac{S}{T_C})}$ [see (\ref{eq:Q})],
identifying $N_0=T_C$ as the thermal noise power spectral density.

\textit{Carnot efficiency.---}
The minimum value of the last relation is
$Q=IT_{C}$, obtained when $S\rightarrow0$, i.e., when the engine works
reversibly with minimal energy consumption. As we invest $Q$ and
gain $\bar E$, the net average energy gain of the engine is
\bea W=\bar E - I T_C = \bar{E}\left(1-T_C/T_H\right), \eea
resulting in the maximal efficiency $\eta_C= 1- T_C/T_H$ \cite{gas}.

\textit{Maximum power efficiency.---} We calculate
$\eta_{M}$ using the Gaussian capacity (\ref{eq:Gauss2}). The power
(\ref{eq:P}) reduces to
\bea P =
T_{H}\ln\left(1+S/T_C\right)-S. \eea
Taking $\frac{\partial P}{\partial S}=0$ yields the energy per pulse
maximizing the power, $S_M=T_H-T_C$. We substitute
this result into (\ref{eq:etaM}) and obtain
$\eta_M= 1 -S_M/\left[  T_H\ln\left(1+\frac{S_M}{T_{C}}\right)\right]$.
In terms of the Carnot efficiency, $\eta_{C}=\frac{S_M}{T_H}$, we find that
\bea \frac{\eta_M}{\eta_C} & = &
\frac{1}{\eta_C}+\frac{1}{\ln\left(1-\eta_C\right)}\nonumber\\
& = &
\frac{1}{2}+\frac{1}{12}\eta_C+\frac{1}{24}\eta_C^{2}+{\cal O}\left(\eta_C^{3}\right).
\eea
%
For $\eta_C\rightarrow 1$, the maximal power efficiency converges to 1.
Note that the coefficient of the second (quadratic) term is smaller than the value
recovered in several spatially-symmetric systems \cite{Katja1}. 


\begin{figure}
\hspace{2mm}
{\hbox{\epsfxsize=70mm \epsffile{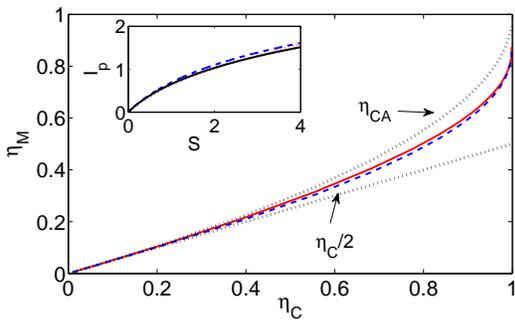}}} \caption{ Maximum power
efficiency for a heat engine of a narrowband bosonic channel
with $\nu=0.2$ (full) and $\nu=2$ (dashed). $T_C=1$ was kept fixed. 
The CA efficiency and half of the Carnot limit are shown as dotted
lines. Inset: Gaussian (dashed line), wideband bosonic (dotted line)
and narrowband bosonic with $\nu=2$ (full line) capacities,
$T_C=1$.}
\label{FigN}
\end{figure}

\textit{Case II: Bosonic Channel.---}
We next send the same {\it classical} information, yet in the form of quantum states,
over a memoryless quantum channel which is a bosonic field (e.g. an electromagnetic radiation): The message
information is encoded into modes of frequency $\nu$ and average
photon number $n_B(\nu)$ \cite{Caves,Vedral}. The signal power is denoted by
$\mathcal S$; noise power is $N= \pi T_C^2/12\hbar$ \cite{Blencowe}.
Under the constraint that the message-ensemble-averaged energy of the channel is fixed, 
the capacity of this channel is given by
$C=\frac{\pi}{6\hbar}(T_e-T_C)$, where $T_e$, the (effective) temperature of the decoder, is defined
through the relation
${\mathcal S}+\frac{\pi T_C^2}{12\hbar}=\frac{\pi T_e^2}{12\hbar}$, leading to
%
$C=\frac{\pi T_C}{6\hbar} \left[ (12\hbar {\mathcal S}/\pi T_C^2 +1)^{1/2}
-1\right]$. 
%
Redefining $I_p=C/B$ and $S={\mathcal S}/B$; $B$ denotes an
effective bandwidth, we obtain
\bea
I_p= \frac{\pi T_C}{6B\hbar}
\left[ \left( \frac{12 \hbar S B}{\pi T_C^2} +1 \right)^{1/2} -1 \right].
\label{eq:Bosonic}
\eea
Remarkably, this capacity is inherently linked to the thermal conductance $\kappa$  of
a ballistic quantum wire \cite{Blencowe}.

\textit{Carnot efficiency.---} We again minimize heat production in
the channel $Q=IS/I_p$, using (\ref{eq:I}) and (\ref{eq:Bosonic}).
The minimum value of this expression is ($\lambda\equiv 6B\hbar/\pi$)
\bea Q_{S\rightarrow0}&=&\lim_{S\rightarrow 0} \frac{ \bar E}{T_H}
\frac{ \lambda S/T_C}{\left(2\lambda S / T_C^2+1\right)^{1/2}-1}
\nonumber\\ &=& \bar ET_C/T_H, \eea
retrieving the Carnot limit, $\eta_C=(\bar E -Q_{S\rightarrow 0})/\bar E=(1-T_C/T_H)$.

\textit{Maximum power efficiency.---}
Maximizing $P$, we obtain $S_M=\frac{ T_C^2}{2\lambda} \left[ \left( \frac{T_H}{T_C}\right)^2-1\right]$.
We plug it into (\ref{eq:etaM}) and exactly resolve {\it only} the linear term
\bea
\eta_M/\eta_C=1/2.
\eea
Interestingly, this result holds in both the high temperature limit,
when the classical Shanon capacity is recovered $I_p=S/T_C$, as well as at
low temperatures when quantum effects become important and the
channel capacity approaches $I_p=\sqrt{\pi S  /3\hbar B}$.

\begin{figure}
\hspace{2mm}
{\hbox{\epsfxsize=70mm \epsffile{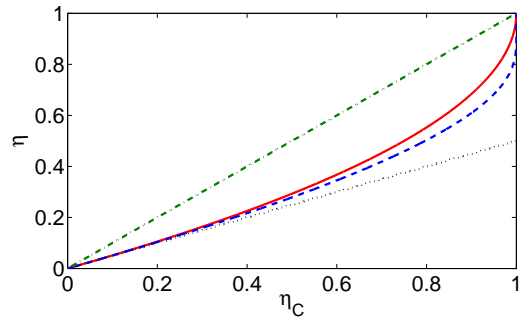}}} \caption{Maximum power
efficiency using different models for the information channel:
Gaussian (\ref{eq:Gauss2}) (dashed line),  wideband Bosonic
(\ref{eq:Bosonic}) (dotted), compared to the Curzon-Ahlborn
efficiency (full) and the Carnot efficiency (dashed-dotted line). }
\label{Figeta}
\end{figure}

We proceed and explore a narrowband photon channel, $B\ll \nu$,
where $\nu$ is the channel central frequency. In this case the noise
power is given by $N=h \nu n_B(\nu,T_C) B$;
$n_B(\nu,T)=[e^{h\nu/T}-1]^{-1}$ is the Bose-Einstein distribution.
Based on the assumption that photon noise is additive, the
(dimensionless) capacity $I_p=C/B$ fulfills \cite{EBook}
%
\bea
&&I_p=\ln\left[ 1+ \frac{S}{h\nu} (1-e^{-h\nu/T_C}) \right] +[S/h\nu+n_B(\nu,T_C)]
\nonumber\\
&&\times
\ln\left[ 1+ \frac{h\nu }{S+h\nu n_B(\nu,T_C) }\right] -n_B(\nu,T_C)h\nu/T_C.
\eea
Since analytical results are cumbersome, we acquire $S_M$ numerically,
seeking the intersect of $1/T_H$ with $\partial I_p(S)/\partial S$, see Eq. (\ref{eq:P}).
The results, displayed in Fig. \ref{FigN}, clearly demonstrate that
$\eta_M$ is bounded by $\eta_{CA}$ from above, and that $\eta_M=\eta_C/2$ in the lowest order of $\Delta T/T$.
Interestingly, the results are almost independent of the channel central frequency $\nu$.
The inset displays channel capacities for the three models considered: Gaussian, wideband  and narrowband
bosonic channels.
Figure \ref{Figeta} further shows the maximum power efficiency $\eta_M$ for
the classical and quantum wideband channels, as a function of $\eta_C$,
compared to the CA and the Carnot efficiencies.
We find that the CA value is an upper bound for $\eta_M$.


\textit{Case III: Generic Channel.---} Motivated by the classical-Gaussian
channel and the quantum models, we consider next a general form for a channel capacity,
\bea I_p= \alpha x +\beta x^2+\gamma x^3+ {\cal O}(x^4).
\label{eq:IpG}
\eea
Here $x=S/T_C$ is the signal power over noise,  and the parameters
$\alpha$, $\beta$ and $\gamma$ may depend on $T_C$.
These coefficients are constricted so as to produce positive capacity, $I_p\geq0$.
Assuming $I_p$ is a concave function, valid for the cases investigated above, the inequality
$3\gamma S <-\beta T_C$ must be satisfied. Since $S>0$ and $\gamma>0$ [so as $I_p(S\gg0)>0$],
we conclude that $\beta$ must be negative.
In particular, for the Gaussian channel (\ref{eq:Gauss2}) a power expansion gives
$\alpha=1$, $\beta=-1$ and $\gamma=4/3$, while for the bosonic wideband quantum channel we
get $\alpha=1$, $\beta=-3\hbar/\pi T_C$ and $\gamma=18\hbar^2/\pi^2T_C^2$.
Using the generic form (\ref{eq:IpG}),
the maximal efficiency is obtained by minimizing heat production, $Q=IS/I_p$, resulting in
\bea
Q(S\rightarrow 0)= I T_C/\alpha; \,\,\,\ \eta= \tilde \eta_C/\alpha,  
\label{eq:etaCGen}
\eea
where $\tilde \eta_C=\eta_C+\alpha-1$. Thus,
in order to reach the Carnot limit one must require that $\alpha=1$.
Next, still using the form (\ref{eq:IpG}) we maximize $P$ and obtain the pulse power
%
$S_M=\frac{-\beta T_C}{3\gamma} \left[ 1 -\left(1-\frac{3\gamma\tilde\eta_C}{\beta^2} \right)^{1/2}\right]$.
%
Plugging this value into (\ref{eq:etaM}) leads to
\bea
&&\eta_M
 =\frac{\tilde \eta_C}{2\alpha} +\frac{\tilde \eta_C^2}{4} \left(\frac{1}{\alpha^2} - \frac{\gamma}{2\beta^2\alpha}
\right) +{\cal O}(\tilde \eta_C^3).
\label{eq:etaMG}
\eea
Imposing $\alpha=1$ results in
\bea
&&\eta_M/\eta_C= \frac{1}{2}+ \frac{\eta_C}{4}(1-\gamma/2\beta^2) +{\cal O}(\eta_C^2).
\label{eq:etaMG2}
\eea
%
The following important observations can be made:
(i) The Carnot limit $\eta_C=(T_H-T_C)/T_H$ and the efficiency at
maximum power $\eta_M= \eta_C/2$ (at first order in $\eta_C$) are
non-subordinate elemental results in thermodynamics:
Demanding that $\alpha=1$ for channels working at maximal capacity,
independently produces both the Carnot efficiency,
and the universal linear coefficient in Eq. (\ref{eq:etaMG2}).
(ii) A universal relation between the two efficiencies stands still
even {\it below} the upper bound: Assuming that $\alpha<1$ accounts
for loss mechanisms in the channel, operating below its full
capacity, the maximal efficiency of the engine is given by
$\tilde \eta/\alpha$, 
while the maximal power efficiency (in linear response)
is exactly half this value, $\tilde \eta/2\alpha$. 
(iii) The first (linear) term in (\ref{eq:etaMG}) is independent of the
details of the channel model, $\beta$ and $\gamma$.
%
(iv) Finally, the (lack of) universality in the quadratic term of $\eta_M$ has been a source of debates \cite{Katja1}.
Our analysis exposes that this coefficient depends on the details of the channel,
i.e., on the relations between the nonlinear terms. In
particular, if $\gamma=2\beta^2$ the quadratic term 
diminishes, in agreement with the quantum (wideband) bosonic channel. When
$4\beta^2=3\gamma$, it reduces to $\eta_C^2/12$, as in
the classical Gaussian channel. For $\gamma=\beta^2$ the quadratic
term becomes $\eta_C^2/8$, recovering the CA result (\ref{eq:etaCA}) and other symmetric models \cite{Katja1}.

\textit{Summary.---} We have designed here a minimal model of a heat
engine (or a Maxwell's demon), attributing energy dissipation within
the machine to irreversible loss of energy within the engine
communication channels.
Analyzing both classical and quantum information channels
we have demonstrated that our model satisfies the Carnot limit for reversible operation mode.
Independently, it leads to the universal linear term in the maximum power efficiency,
exhibiting the existence of fundamental laws for finite-time thermodynamic processes.
While the universality of the maximal power efficiency in the linear
response regime has been explained based on the Onsager's symmetry
\cite{broeck}, our model establishes this universality developing a
distinct, information based picture.
We expect this prototype model to be immensely useful for studying
other basic concepts in thermodynamics. For example, the engine
described here could serve as a cooling device, or the performance
could be analyzed for a subsystem away from the Boltzmann
distribution. One could also advance the model beyond the strong
coupling limit, $W \propto \bar E$ \cite{Strong}, or consider other
sources of dissipation. Finally, it is of great interest to further
include quantum effects, considering transmission of quantum
information. 


\noindent \emph{Acknowledgments.} This research has been supported
by NSERC.


\begin{thebibliography}{32}
\bibitem{units}
Standardly, the units of the channel capacity $C$ are nats/sec, and it is given in
terms of the signal power $\mathcal S$ and the noise power $N$.
Here $I_p$ is a dimensionless channel capacity, given in terms of the
signal and noise averaged energy, $S$ and $N_0$, respectively.
For simplicity, we use 2 pulses as the basic unit for counting information transfer in a channel
\cite{EBook}. This choice does not affect our results.
\bibitem{gas}
The model could be also formulated using a one-molecule gas. In an engine
cycle it is reversibly compressed to the volume $V/2$ at
temperature $T_C$, then expanded back to $V$ at $T_H$. In the
compression phase the information $I=\ln(2)$ is obtained with an
energy cost $Q=T_C\ln(2)$. The information is used in the expansion
phase to gain $\bar E =T_H\ln(2)$, yielding the Carnot efficiency.
\bibitem{Strong}
We refer to this engine as "strongly coupled" since
the work attained $\bar E-Q\propto I$ is proportional to the energy pumped,
$\bar E=IT_H$ (\ref{eq:I}).
\end{thebibliography}
\end{document}